\documentclass[twocolumn,showpacs,floatfix,prl]{revtex4}%
\usepackage{graphicx}%
\usepackage{amsmath}%
\usepackage{amsfonts}%
\usepackage{amssymb}
\usepackage{bm}

\def\s{{\sigma}}

\def\k{{ {\bm k} }}

\def\q{{ {\bm q} }}
\def\Q{{ {\bm Q} }}
\def\0{{ {\bm 0} }}

\def\w{{\omega}}
\def\a{{\alpha}}
\def\b{{\beta}}

\def\g{{\gamma}}

\allowdisplaybreaks[4]

\begin{document}
\title{
Spin Triplet Superconductivity in Sr$_2$RuO$_4$ due to 
Orbital and Spin Fluctuations: \\
Analysis by Two-Dimensional Renormalization Group Theory
}
\author{
Masahisa \textsc{Tsuchiizu}$^{1}$, 
Youichi \textsc{Yamakawa}$^{1}$, 
Yusuke \textsc{Ohno}$^{1}$, 
Seiichiro \textsc{Onari}$^{2}$, and
Hiroshi \textsc{Kontani}$^{1}$
}

\date{\today }

\begin{abstract}
We study the mechanism of the triplet superconductivity
in Sr$_2$RuO$_4$ based on the multiorbital Hubbard model.
The electronic states are studied 
using the renormalization group method.
Thanks to the vertex correction (VC) for the susceptibility,
which is dropped in the mean-field-level approximations,
strong orbital and spin fluctuations at $\Q\approx(2\pi/3,2\pi/3)$
emerge in the quasi one-dimensional Fermi surfaces
composed of $d_{xz}$ and $d_{yz}$ orbitals.
Due to the cooperation of both fluctuations,
we obtain the triplet superconductivity in the $E_u$ representation,
in which the superconducting gap is given by the 
linear combination of 
$(\Delta_x(\k),\Delta_y(\k))\sim (\sin 3k_x,\sin 3k_y)$.
These results are confirmed by a diagrammatic calculation
called the self-consistent VC method.

\end{abstract}

\address{
$^1$ Department of Physics, Nagoya University,
Furo-cho, Nagoya 464-8602, Japan. 
\\
$^2$ Department of Applied Physics, Nagoya University,
Furo-cho, Nagoya 464-8603, Japan. 
}
 

\sloppy

\maketitle


Sr$_2$RuO$_4$ is an unconventional superconductor 
with the transition temperature $T_{\rm c}=1.5$K 
\cite{Maeno,Maeno2,Sigrist-Rev}.
This material has been attracting great attention 
since the spin triplet superconductivity (TSC) is indicated by 
the NMR measurements \cite{Ishida}.
From the early stage, the chiral $p$-wave ($p_x+ip_y$) TSC,
which is analogous of the A-phase of the superfluid $^3$He,
had been predicted \cite{Rice}.
However, in contrast to the paramagnon mechanism in $^3$He,
no ferro-magnetic fluctuations are observed in Sr$_2$RuO$_4$.
Instead, strong antiferro-magnetic (AFM) fluctuations with 
$\Q\approx (2\pi/3,2\pi/3)$
are observed by neutron scattering spectroscopy \cite{neutron}.
Since the AFM fluctuations give the 
spin singlet superconductivity (SSC) in usual, 
the mechanism of the TSC in Sr$_2$RuO$_4$
has been a long-standing problem in strongly correlated electron systems.

\begin{figure}[!htb]
\includegraphics[width=.99\linewidth]{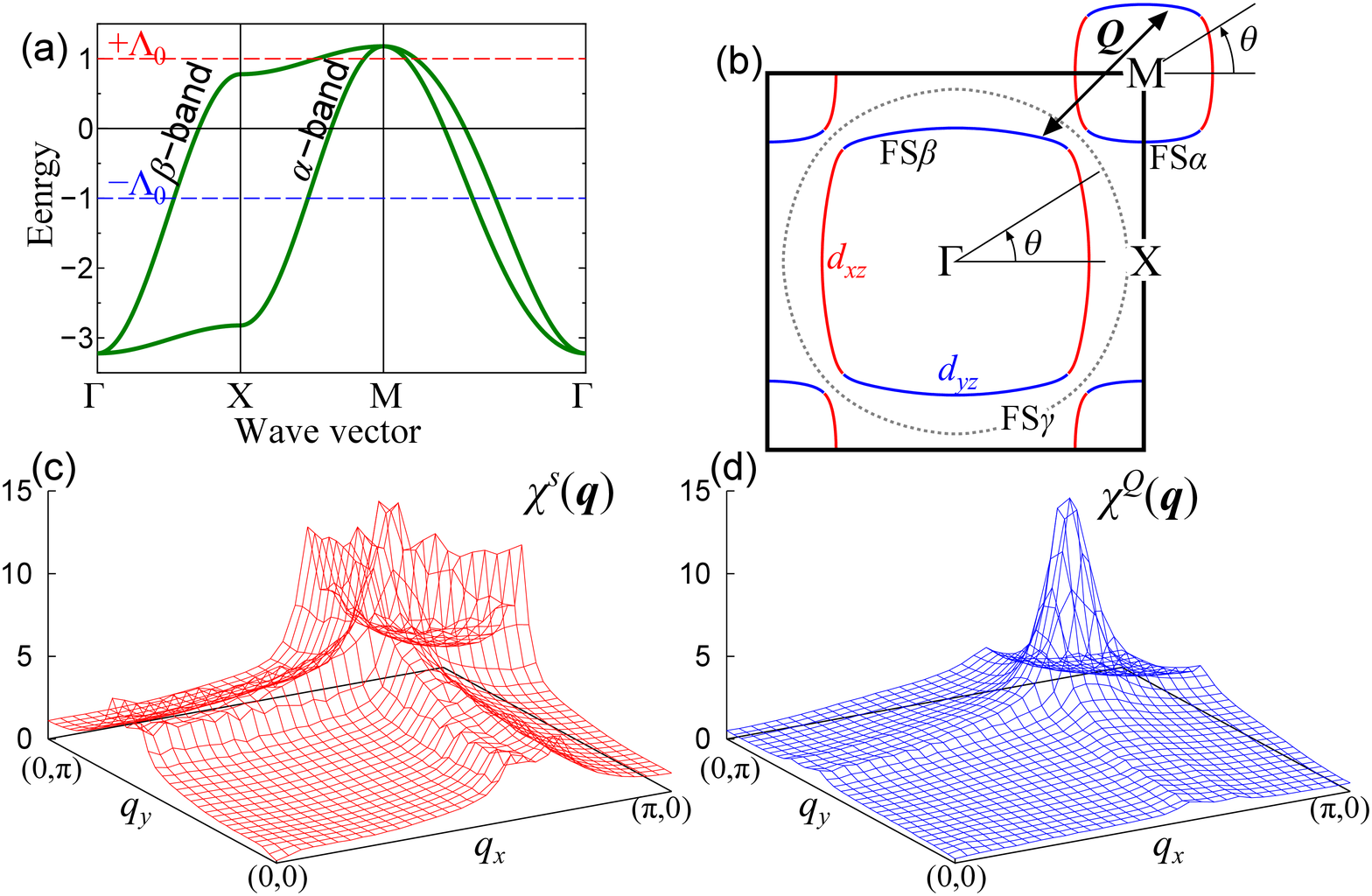}
\caption{(color online)
(a) Bandstructure and (b) FSs of the two-orbital model.
$\Q\approx (2\pi/3,2\pi/3)$ is the nesting vector. 
The FS$\gamma$ of Sr$_2$RuO$_4$ is shown by dotted line.
(c) $\chi^s(\q)$ and (d) $\chi^Q(\q)$ of the q1D-band model
obtained by the RG+cRPA method ($\Lambda_0=1$) 
for $U=3.5$, $J/U=0.035$ and $T=0.02$.
}
\label{fig:FS}
\end{figure}

Figures \ref{fig:FS} (a) and (b) show the bandstructure 
and the Fermi surfaces (FSs) of Sr$_2$RuO$_4$:
The quasi-one-dimensional (q1D) FSs, FS$\a$ and FS$\b$, are composed of 
($d_{xz}$, $d_{yz}$)-orbitals, and the nesting of these q1D FSs
is the origin of the AFM fluctuations at $\Q\approx(2\pi/3,2\pi/3)$.
The two-dimensional (2D) FS, FS$\g$, is composed of only $d_{xy}$-orbital.
If the spin-orbit interaction (SOI) is neglected,
the ($\a,\b$)-bands and $\g$-band are 
coupled only via the electron-electron correlation.
Therefore, the superconductivity would be realized mainly in either 
the q1D bands ($|\Delta_{\a,\b}|\gg |\Delta_\g|$) or 
the 2D band ($|\Delta_{\a,\b}|\ll |\Delta_\g|$).

The mechanisms of the TSC originating mainly from the 2D band
had been proposed in Refs. \cite{Nomura,Wang,Arita,Miyake}:
Nomura and Yamada studied the TSC state using the
perturbation theory \cite{Nomura},
which is the natural development of the 
Kohn-Luttinger mechanism \cite{KL}.
Recently, a three-orbital Hubbard model had been studied
using a 2D renormalization group (RG) method \cite{Wang}.
They obtained the $p$-wave gap on the FS$\g$ accompanied by the  
development of spin fluctuations at $\q=(0.19\pi,0.19\pi)$.
Also, charge-fluctuation-mediated TSC was discussed 
by introducing the inter-site Coulomb interaction \cite{Arita}.

On the other hand, one may expect that the 
TSC is closely related to the AFM fluctuations in the q1D FSs at $\q\sim\Q$.
The TSC originating from the q1D FSs had been discussed
by applying the perturbation theory \cite{Kivelson}
and random-phase-approximation (RPA) \cite{Takimoto,Ogata}.
Takimoto discussed the orbital-fluctuation-mediated TSC
using the RPA under the condition $U'>U$,
where $U$ ($U'$) is the intra-orbital (inter-orbital) Coulomb interaction
\cite{Takimoto}.
However, in the RPA, the SSC is obtained
under the realistic condition $U\ge U'$ due to strong AFM fluctuations.
The TSC due to ferro-charge fluctuations was also discussed \cite{Kohmoto}.
When the spin fluctuation is Ising-like,
the TSC may be favored since the pairing 
interaction for the SSC is reduced \cite{Ogata}.
In these studies, however, it is difficult to obtain the TSC
based on the realistic multiorbital Hubbard model,
under the existence of strong AFM fluctuations as in Sr$_2$RuO$_4$.

To find out the origin of the TSC in Sr$_2$RuO$_4$,
many experimental efforts have been devoted to determine the gap structure,
such as the tunnel junction \cite{Yada},
ARPES, and quasiparticle interference measurements.
Recently, large superconducting gap with $2|\Delta|\approx 5T_{\rm c}$
was observed by the scanning tunneling microscopy measurements
\cite{Firmo}.
The observed large gap would be that on the q1D FSs, 
since the tunneling will be dominated by the  
$(d_{xz},d_{yz})$-orbitals that stand along the $z$-axis,
as clarified in the double-layer compound Sr$_3$Ru$_2$O$_7$ 
\cite{Lee-327}.
Therefore, it is an important challenge
to establish the theory of the TSC based on the q1D-band Hubbard model,
by applying an advanced theoretical method.

In this paper,
we study the mechanism of the TSC 
in Sr$_2$RuO$_4$ based on the realistic $(U>U')$ two-orbital Hubbard model.
The electronic states are studied using the 2D RG method
developed in Ref. \cite{Tsuchiizu}.
Thanks to the vertex correction (VC) for the susceptibility dropped in the RPA,
strong orbital and spin fluctuations at $\Q\approx(2\pi/3,2\pi/3)$
emerge in the q1D bands \cite{Arakawa}.
We propose that the $E_u$-type TSC is realized 
by the cooperation of strong orbital and spin fluctuations in Sr$_2$RuO$_4$.

In this paper, we study the two-orbital Hubbard model,
which describes the quasi-1D FSs of Sr$_2$RuO$_4$.
The kinetic term is given by
$H_0=\sum_{\k,\s}\sum_{l,m}^{1,2}\xi_\k^{l,m}c_{\k,l,\s}^\dagger c_{\k,m,\s}$,
%
where the orbital indices $l,m=1$ and $2$ refer to 
$d_{xz}$- and $d_{yz}$-orbitals, respectively.
In the present model,
$\xi_\k^{1,1}=-2t\cos k_x-2t_{\rm nn}\cos k_y$,
$\xi_\k^{2,2}=-2t\cos k_y-2t_{\rm nn}\cos k_x$, and
$\xi_\k^{1,2}=4t'\sin k_x \sin k_y$.
Hereafter, we set $(t,t_{\rm nn}, t')=(1,0.1,0.1)$, and 
fix the filling as $n=4\cdot (2/3)=2.67$,
which corresponds to the filling of the q1D FSs of Sr$_2$RuO$_4$.
We also introduce the on-site Coulomb interactions $U$, $U'$, 
and put the exchange and Hund's couplings $J=J'=(U-U')/2$ 
throughout the paper.

Here, we analyze this model by applying the RG combined with the 
constrained RPA (RG+cRPA) \cite{Tsuchiizu}.
This method is very powerful to calculate the 
higher-order many-body effects systematically and in an unbiased way.
In the RG+cRPA method, we divide the lower-energy region  ($|E|<\Lambda_0$) 
of the Brillouin zone into $N$ patches as done in 
Refs.\cite{Metzner,Honerkamp,RG-Rev,DHLee} and perform the RG analysis. 
The contributions from the higher-energy region ($|E|>\Lambda_0$)
are calculated by the cRPA method with high numerical accuracy, 
and incorporated into the initial vertex functions \cite{Tsuchiizu}.
(The conventional patch-RG method \cite{Metzner,Honerkamp,RG-Rev,DHLee}
is recovered when $\Lambda_0>W_{\rm band}$.)
Although the initial vertex functions are very small,
they play decisive roles for the fixed point of the RG flow.

We use $N=64$ (32 patches for each FS) in the present study, 
and it is verified that the results of $N=128$ are almost unchanged.
First, we calculate the susceptibilities using the RG+cRPA:
The charge (spin) susceptibility is given by
$\displaystyle \chi_{l,l';m,m'}^{c(s)}(q)= 
\int_0^\beta d\tau
\frac12 \langle A_{l,l'}^{c(s)}(\q,\tau)A_{m',m}^{c(s)}(-\q,0)\rangle e^{i\w_l\tau}$,
where $A_{l,l'}^{c(s)}(\q)=\sum_{\k} (c^\dagger_{\k,l',\uparrow}c_{\k+\q,l,\uparrow}
+(-)c^\dagger_{\k,l',\downarrow}c_{\k+\q,l,\downarrow})$,
$q=(\q,\w_l)$, and $l,l',m,m'$ are $d$ orbitals.
The quadrupole susceptibility with respect to
$O_{x^2-y^2}=n_{xz}-n_{yz}$ is given as
$\chi^Q(\q)= \sum_{l,m}(-1)^{l+m}\chi_{l,l;m,m}^{c}(\q)$.
Figures \ref{fig:FS} (c) and (d) show the obtained 
$\chi^s(\q)= \sum_{l,m}\chi_{l,l;m,m}^{s}(\q)$
and $\chi^Q(\q)$, respectively,
by the RG+cRPA method ($\Lambda_0=1$)
for $U=3.5$ and $J/U=0.035$ at $T=0.02$.
Both susceptibilities have the peak at $\Q\approx(2\pi/3,2\pi/3)$,
which is the nesting vector of the present FSs.
The shape of $\chi^s(\q)$ is essentially equivalent to 
that of the RPA, by putting  $U=2.2$ and $J/U=0.035$.
However,  $\chi^Q(\q)$ in the RPA is quite small when $J>0$
\cite{Onari-SCVC,Ohno-SCVC}.
Therefore, the enhancement of $\chi^Q(\q)$ in Fig. \ref{fig:FS} (d)
originates from the many-body effect beyond the RPA.
The natural candidate is the Aslamazov-Larkin (AL) type
VC for $\chi^Q(q)$, $X^c(q)$,
whose analytic expression is given in Ref. \cite{Onari-SCVC}.
Since $X^c(q) \sim U^4 T\sum_k \Lambda_{\rm AL}(q;k)^2 \chi^s(k)\chi^s(k+q)$
for simplicity, $X^c(q)$ takes large value at $\q=\0$ and $2\Q$
when $\chi^s(\k)$ is large at $\k=\Q$.
$\Lambda_{\rm AL}(q;k)$ is the three-point vertex
composed of three Green functions \cite{Onari-SCVC}.
In the present model, $2\Q\approx\Q$ in the first Brillouin zone.
Thus, with the aid of the VC and the nesting of the FSs,
the enhancement of $\chi^Q(\Q)$ in Fig. \ref{fig:FS} (d) is realized.

\begin{figure}[!htb]
\includegraphics[width=.99\linewidth]{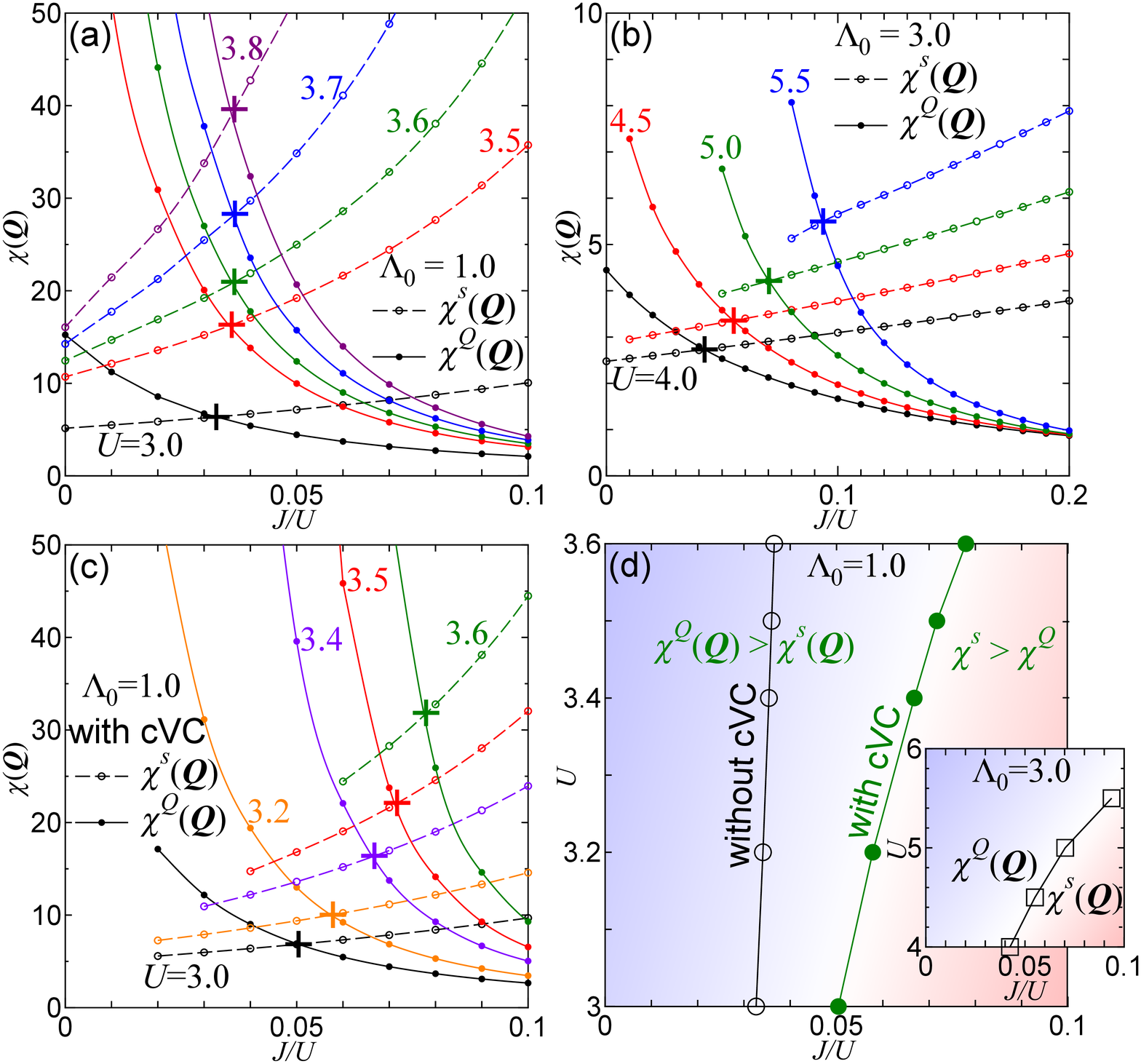}
\caption{(color online)
$\chi^s(\Q)$ and $\chi^Q(\Q)$ as functions of $J/U$
given by the RG+cRPA method for (a) $\Lambda_0=1$ ($U=3.0\sim3.8$)
and (b) $\Lambda_0=3$ ($U=4.0\sim5.5$).
(c) $\chi^s(\Q)$ and $\chi^Q(\Q)$ for $\Lambda_0=1$,
by including the constrained VC (cVC).
(d) Obtained phase diagram for $\Lambda_0=1$
and $\Lambda_0=3$ (inset).
}
\label{fig:chiQ}
\end{figure}

Figure \ref{fig:chiQ} (a) shows $\chi^s(\Q)$ and $\chi^Q(\Q)$ 
as functions of $J/U$ at $T=0.02$, 
obtained by the RG+cRPA method with $\Lambda_0=1$.
For each value of $U$,
$\chi^Q(\Q)$ ($\chi^s(\Q)$) decreases (increases) with $J/U$,
and they are equal at $(J/U)_{c}\sim0.035$.
We stress that $(J/U)_{c}$ is negative in the RPA
since the VC is totally dropped.
In the case of $\Lambda_0=3$ shown in Fig. \ref{fig:chiQ} (b), 
the value of $(J/U)_{c}$ 
increases to $\sim0.08$ at $U\sim5$, indicating that importance 
of the VC due to higher energy region.
To check this expectation, we include the 
constrained AL term (cVC) in addition to the cRPA \cite{Tsuchiizu}.
The obtained results are shown in Fig. \ref{fig:chiQ} (c).
It is verified that $(J/U)_{c}$ increases to $0.08$ at $U=3.7$.
($\chi^Q(\Q)$ in Fig. \ref{fig:chiQ} (c) is approximately given by 
shifting $\chi^Q(\Q)$ in Fig. \ref{fig:chiQ} (a) 
horizontally by $+0.02\sim+0.05$.)
The values of $(J/U)_{c}$ obtained by Figs. \ref{fig:chiQ} (a)-(c) 
are summarized in Fig. \ref{fig:chiQ} (d).
Note that $(J/U)_{c}\sim0.1$ ($\sim0.15$) in the SC-VC$_{(\Sigma)}$ method
\cite{Onari-SCVC,Onari-Hdoped}.

\begin{figure}[!htb]
\includegraphics[width=.99\linewidth]{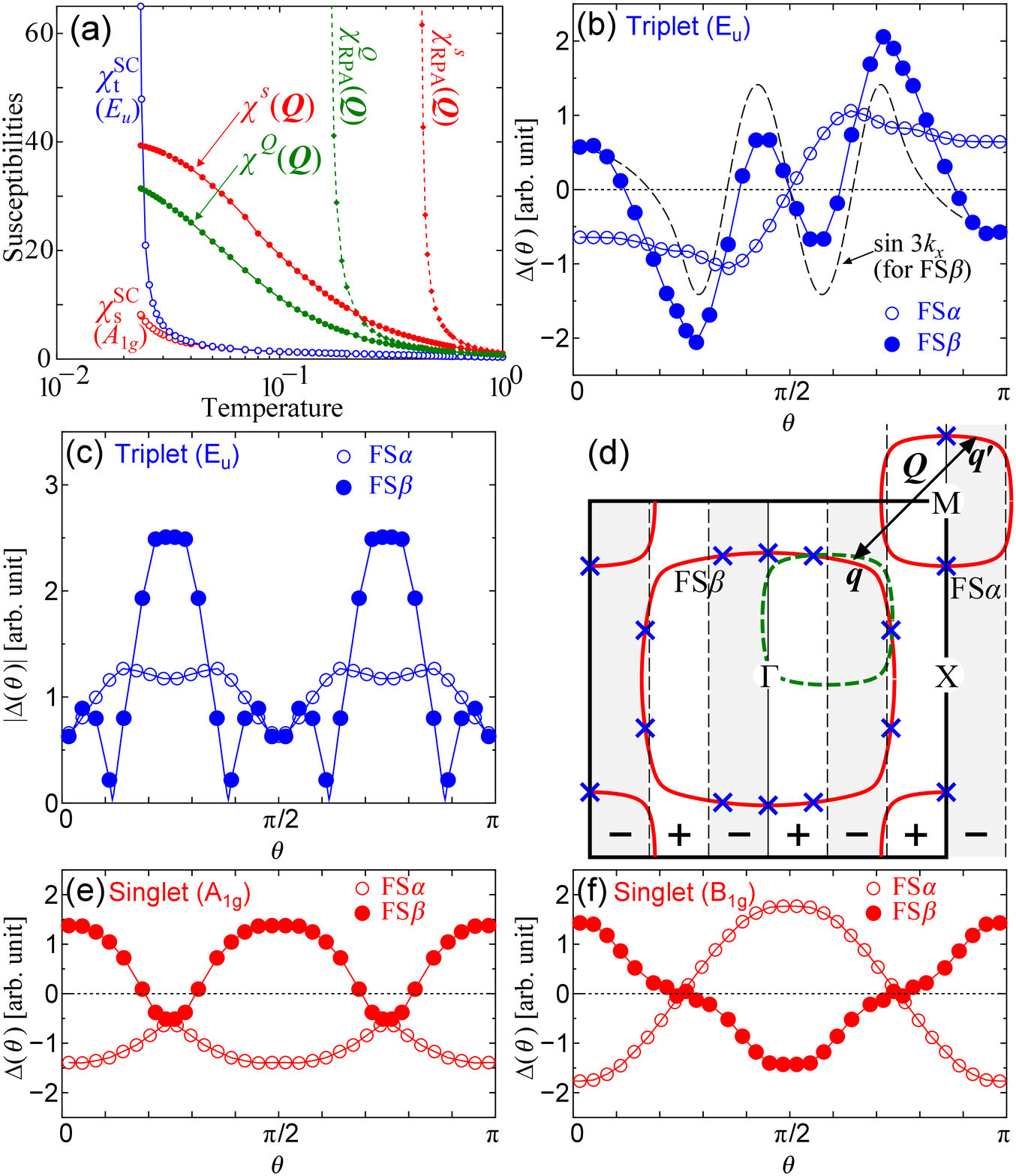}
\caption{(color online)
(a) $T$-dependences of $\chi^s(\Q)$, $\chi^Q(\Q)$,
$\chi_{\rm s}^{\rm SC}$ and $\chi_{\rm t}^{\rm SC}$ for $U=3.8$ and $J/U=0.04$
($\Lambda_0=1$).
(b) $E_u$ gap functions on FS$\mu$, $\Delta_x^\mu(\theta)$
($\mu=\a,\b$) obtained by the RG.
The relation $\Delta_x^\b \propto \sin 3k_x$ holds approximately.
$N=128$ patches are used.
(c) The magnitude of the chiral (or helical) gap state
$|\Delta^\mu|= \sqrt{(\Delta_x^\mu)^2+(\Delta_y^\mu)^2}$.
(d) Schematic explanation for the $\sin 3k_x$-type TSC due to 
orbital+spin fluctuations at $\q=\Q$.
Solid lines (broken lines) are the necessary (accidental) nodes.
The positions of nodes ($\Delta_x^\mu=0$) in (b) are shown by crosses.
(e) $A_{1g}$ and (f) $B_{1g}$ SSC gap functions.
}
\label{fig:SC}
\end{figure}

Although the value of $(J/U)_{c}$ is underestimated at $\Lambda_0=1$,
the obtained $\chi^s(\q)$ and $\chi^Q(\q)$ at $\Lambda_0=1$ is reliable,
since the higher-energy processes can be calculated
with high numerical accuracy
 \cite{Tsuchiizu}.
Hereafter, we perform the RG+cRPA method with $\Lambda_0=1$,
by using smaller $J/U\ (\sim0.04)$ to 
compensate for the absence of the higher-energy VCs.
Figure \ref{fig:SC} (a) shows the $T$-dependences of
$\chi^s(\Q)$ and $\chi^Q(\Q)$ given by the RG+cRPA method
($\Lambda_0=1$) for $U=3.8$ and $J/U=0.04$:
Both of them are strongly renormalized from the RPA results.
In the RPA, $\chi^s_{\rm RPA}(\Q)$ diverges at $T\approx 0.4$,
at which $\chi^Q_{\rm RPA}(\Q)$ remains very small.
In highly contrast, in the RG+cRPA method,
the relation $\chi^s(\Q) \approx \chi^Q(\Q)$ holds for wide temperature range.

We also calculate the TSC and SSC susceptibilities
using the RG+cRPA method:
\begin{eqnarray}
\chi_{\rm t(s)}^{\rm SC}=
\frac12 \int_0^\beta d\tau\langle B_{\rm t(s)}^\dagger(\tau)B_{\rm t(s)}(0)\rangle,
\label{eqn:chiSC}
\end{eqnarray}
where $B_{\rm t(s)} = 
\sum_{\q,\mu}\Delta_{\rm t(s)}^\mu(\q) c_{\q,\mu,\uparrow} c_{-\q,\mu,\uparrow(\downarrow)}$.
$\mu=\a,\b$ is the band index, and
$\Delta_{\rm t(s)}^\mu(\q)$ is the odd (even) parity gap function.
The obtained $\chi_{\rm t(s)}^{\rm SC}$ is shown in Fig. \ref{fig:SC} (a),
by optimizing the functional form of $\Delta_{\rm t(s)}^\mu(\q)$ numerically
\cite{comment-d}.
Since $\chi_{\rm t(s)}^{\rm SC}$ diverges at $T=T_{\rm c}$, 
the strong development of $\chi_{\rm t}^{\rm SC}$ at $T\approx 0.02$
means that the TSC is realized.
This TSC state belongs to the two-dimensional $E_u$-representation,
$(\Delta_{x}^\mu(\q), \Delta_{y}^\mu(\q))$.
The obtained $\Delta_{x}^\mu$ on the FSs when $\chi^{\rm SC}_{\rm t}\sim60$
are shown in Fig. \ref{fig:SC} (b), where $\theta$ is the 
angle of the Fermi momentum shown in Fig. \ref{fig:FS} (b).
The necessary nodes $\Delta_{x(y)}^\mu=0$ 
are on the lines $q_{x(y)}=0,\pm\pi$.
Very similar TSC gap is obtained for $J/U\lesssim0.08$
by taking the cVC into account with $\Lambda_0=1$.
Below $T_{\rm c}$, the BCS theory tells that the
chiral or helical gap state with the gap amplitude
$|\Delta^\mu|= \sqrt{(\Delta_x^\mu)^2+(\Delta_y^\mu)^2}$,
which is shown in Fig. \ref{fig:SC} (c),
is realized to gain the condensation energy.

To understand why the TSC state is obtained,
it is useful to analyze the linearized gap equation:
\begin{eqnarray}
\lambda_{\rm a}^E {\bar \Delta}_{\rm a}^\mu(\q)
&=& -\sum_{\mu'}^{\a,\b} \int_{\rm FS\mu'}\frac{d\q'}{v_{\q'}^{\mu'}}
V_{\rm a}^{\mu,\mu'}(\q,\q') {\bar \Delta}_{\rm a}^{\mu'}(\q')
 \nonumber\\
& &\times {\rm ln}(1.13\w_c/T),
\label{eqn:GapEq}
\end{eqnarray}
where a = t or s.
$\lambda^E_{\rm a}$ is the eigenvalue, 
$V_{\rm a}^{\mu,\mu'}(\q,\q')$ is the pairing interaction, and
$\w_c$ is the cut-off energy of the interaction.
As shown in Fig. \ref{fig:FS} (b), 
the inter-band interaction ($\mu=\a$, $\mu'=\beta$)
with $\q-\q'=\Q$ is approximately given by the intra-orbital interaction
given as
\begin{eqnarray}
{V}_{\rm a}^l(\q;\q')
&=& b_{\rm a}\frac{U^2}{2} |{\Lambda}_l^s(\q;\q')|^2
{\chi}_l^s(\q-\q')
\nonumber \\
& &+ c_{\rm a}\frac{U^2}{2} |{\Lambda}_l^c(\q;\q')|^2
{\chi}_l^c(\q-\q') ,
 \label{eqn:Vts} 
\end{eqnarray}
where $(b_{\rm t},c_{\rm t})=(-1,-1)$ and $(b_{\rm s},c_{\rm s})=(3,-1)$,
and ${\chi}_l^{s,c}(\Q)\equiv {\chi}_{l,l;l,l}^{s,c}(\Q)$.
(Note that ${\chi}_l^s(\Q)\approx{\chi}^s(\Q)/2$ 
and ${\chi}_l^c(\Q)\approx{\chi}^Q(\Q)/4$,
since ${\chi}_{l}^{s}(\Q)\gg {\chi}_{1,1;2,2}^{s}(\Q)$
and ${\chi}_{l}^{c}(\Q)\approx -{\chi}_{1,1;2,2}^{c}(\Q)$
near the critical point \cite{Ohno-SCVC}.)
${\Lambda}_l^{s,c}$ is the VC for the gap equation,
which we call $\Delta$-VC in Ref. \cite{Onari-Hdoped}.
The AL-type diagram for the charge channel is given by
${\Lambda}_l^{c}(\q;\q') \sim
1+T\sum_k \Lambda_{\rm AL}(q-q';k)G(k) \chi^s(k+q)\chi^s(k-q')$,
which is strongly enlarged for $\q-\q'\approx \Q$,
and the orbital-fluctuation-mediated pairing is favored
\cite{Onari-SCVC,Onari-Hdoped}. 
The merit of the RG+cRPA method is that the
the  AL-type $\Delta$-VC is automatically produced
in calculating the pairing susceptibility in Eq. (\ref{eqn:chiSC}).

In the RPA with $J>0$,
the TSC cannot be achieved because of the relation 
$\chi_l^s(\Q) \gg \chi_l^c(\Q)$ and $\Lambda^{c,s}=1$ in the RPA:
In this case, spin-fluctuation-mediated SSC is obtained since
$|{V}_{\rm s}^l|= (U^2/2)\{3|\Lambda_l^s|^2\chi_l^s-|\Lambda_l^c|^2\chi_l^c\}$ 
is three times larger than 
$|{V}_{\rm t}^l|= (U^2/2)\{|\Lambda_l^s|^2\chi_l^s+|\Lambda_l^c|^2\chi_l^c\}$.
In the present RG+cRPA method, in contrast, the relationship
$\chi_l^s(\Q) \sim \chi_l^c(\Q)$ is realized, and therefore
the triplet interaction $|{V}_{\rm t}^l|$
can be larger than $|{V}_{\rm s}^l|$.
Using Fig. \ref{fig:SC} (d), we explain the gap structure of the 
TSC state induced by orbital+spin fluctuations at $\q\approx\Q$.
In addition to the necessary nodes shown by solid lines, 
accidental nodal lines appear around
$k_x\approx \pm\pi/3$ and $k_x\approx \pm2\pi/3$:
The reason is that $\Delta_{x}^\a(\q)$ and $\Delta_{x}^\b(\q')$ 
tend to have the same sign for $\q-\q'\approx \Q$
due to large attractive interaction by $V_{\rm t}^l(\q;\q')$.
For this reason, the relation 
$\Delta_{x}^\b(\q) \sim \sin 3k_x$ in Fig. \ref{fig:SC} (b)
is satisfied in the $E_u$-type TSC state.

In Fig. \ref{fig:SC} (a),
$\chi_{\rm s}^{\rm SC}$ also develops at low temperatures:
Figures  \ref{fig:SC} (e) and (f) show the obtained 
$A_{1g}$ and $B_{1g}$ SSC gap structures, which give
the first and the second largest $\chi_{\rm s}^{\rm SC}$'s.
Both SSC states with sign reversal are mainly caused by 
spin fluctuations, and $A_{1g}$ state is slightly stabilized 
by the orbital fluctuations.
The $A_{1g}$ state in Fig. \ref{fig:SC} (e) dominates the TSC state
when $\chi^s(\Q)\gg\chi^Q(\Q)$, which is realized for $J/U\gtrsim0.05$
in Fig. \ref{fig:chiQ} (a).

\begin{figure}[!htb]
\includegraphics[width=.99\linewidth]{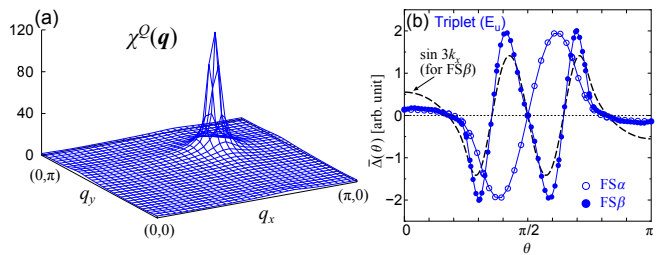}
\caption{(color online)
Numerical results obtained by the SC-VC method:
(a) $\chi^Q(\q)$ and 
(b) TSC gap function ${\bar \Delta}_{x}^\mu$.
}
\label{fig:SCVC}
\end{figure}

To verify the reliability of the results given by the RG+cRPA method,
we also study the present model using the SC-VC method \cite{Onari-SCVC}.
Figure \ref{fig:SCVC} (a) shows the obtained $\chi^Q(\q)$
for $U=2.33$ and $J/U=0.1$ at $T=0.05$.
Its peak position at $\q\approx\Q$ is consistent with 
the RG+cRPA result in Fig. \ref{fig:FS} (d).
$\chi^Q(\Q)=113$ and $\chi^s(\Q)=46$ in the present calculation.
By taking the self-energy correction into SC-VC method,
the orbital fluctuations will develop even for $J/U \sim 0.15$
\cite{Onari-Hdoped}.
Next, we can study the superconducting state 
by solving the linearized gap equation.
The obtained largest eigenvalue is 
$\lambda_{\rm t}^{\rm SC}=0.495$ ($E_u$ state) and 
$\lambda_{\rm s}^{\rm SC}=0.479$ ($A_{1g}$ state).
The obtained TSC gap function is shown in Fig. \ref{fig:SCVC} (b),
which is essentially similar to the gap structure in Fig. \ref{fig:SC} (b).
Thus, the numerical results of the RG+cRPA method
are confirmed by the diagrammatic approach.

The filling of the q1D bands in Sr$_2$RuO$_4$ is $n=2.8$
according to the band calculation \cite{Wien2k}.
Even in this case, the TSC state with $\Delta_{x(y)}^\b \sim\sin 3k_{x(y)}$
is also obtained, by using both RG+cRPA and SC-VC methods.
The obtained peaks of $\chi^Q(\q)$ and $\chi^s(\q)$ coincide
and shifts to $\q\approx(0.6\pi,0.6\pi)$.

Even in the RPA,
strong orbital fluctuations can be obtained by putting $U'>U$
 \cite{comment}.
The TSC can be realized by
orbital fluctuations as found by Takimoto \cite{Takimoto},
but the fully-gapped $A_{1g}$ state is also a natural candidate.
Within the RPA, the SSC state is obtained for any $J=(U-U')/2$, and
fully-gapped $A_{1g}$ appears for largely negative $J$.
To obtain the TSC within the RPA, 
we have to choose the ratios $U'/U>1$ and $J/U$ independently
to maintain the coexistence of orbital and spin fluctuations.
In contrast, in the RG+cRPA method,
both fluctuations coexist due to the orbital-spin mode-coupling,
and the TSC is obtained for a wide range of parameters 
under the condition $J=(U-U')/2>0$.

When the TSC occurs in the q1D FSs in real compound, 
the superconducting gap on FS$\g$ will be induced from q1D FSs 
(proximity effect), due to weak inter-band electron correlation
in addition to the large SOI of 4$d$-electron.
As for the latter effect,
large orbital mixture between FS$\b$
and FS$\g$ due to the SOI is predicated by the 
first-principle study \cite{Wien2k}.
It is an important future problem to 
study the TSC in three-orbital model for Sr$_2$RuO$_4$,
by taking the SOI into account.
The ${\bm d}$-vector \cite{Ng,Yanase}
and the topological properties of the TSC state
\cite{Read,Matsumoto,Furusaki,Tanaka-Rev}
can be discussed by this study.

In summary,
we proposed the orbital+spin fluctuation-mediated TSC in Sr$_2$RuO$_4$
by analyzing the two-orbital Hubbard model using the RG+cRPA method.
Thanks to the VC neglected in the RPA,
strong orbital and spin fluctuations at $\q\sim\Q$ emerge in the q1D FSs.
The TSC is obtained for $J/U\lesssim0.04$ (0.08) 
without (with) the cVC for $\Lambda_0=1$.
Similar TSC gap structure is obtained by the SC-VC method
for $J/U\lesssim0.1$.
The present work demonstrated that the RG+cRPA method is very powerful 
in the study of various 2D strongly correlated systems,
emergence of orbital/spin order and superconductivity.

\acknowledgements
We are grateful to K. Yamada, Y. Maeno, Y. Matsuda, K. Ishida, T. Takimoto,
and T. Nomura for fruitful discussions.
This study has been supported by Grants-in-Aid for Scientific 
Research from MEXT of Japan.



\begin{thebibliography}{99}
\bibitem{comment-d}
By applying the present method to the single-band square-lattice Hubbard model,
the $d_{x^2-y^2}$-wave SSC gap function is well reproduced
near the half-filling.
\bibitem{comment}
In the five-orbital model for Fe-based superconductors,
the RPA with $U'>U$ cannot reproduce the strong ferro-quadrupole 
susceptibility that triggers the structure transition.
It can be explained by the SC-VC method \cite{Onari-SCVC}.
\end{thebibliography}
\end{document}